# Thickness-dependent magnetic properties in Pt/[Co/Ni]$_n$ multilayers with perpendicular magnetic anisotropy*

Chunjie Yan(晏春杰)[1], Lina Chen(陈丽娜)[1,2,#], Kaiyuan Zhou(周恺元)[1], Liupeng Yang(杨留鹏)[1], Qingwei Fu(付清为)[1], Wenqiang Wang(王文强)[1], Wen-Cheng Yue(岳文诚)[3], Like Liang(梁立克)[1], Zui Tao(陶醉)[1], Jun Du(杜军)[1], Yong-Lei Wang(王永磊)[3] and Ronghua Liu(刘荣华)[1*]

[1]National Laboratory of Solid State Microstructures, School of Physics and Collaborative Innovation Center of Advanced Microstructures, Nanjing University, Nanjing 210093, China

[2]School of Science, Nanjing University of Posts and Telecommunications, Nanjing 210023, China.

[3]School of Electronics Science and Engineering, Nanjing University, Nanjing 210093, China.

We systematically investigated the Ni and Co thickness-dependent perpendicular magnetic anisotropy (PMA) coefficient, magnetic domain structures, and magnetization dynamics of Pt(5 nm)/[Co($t_{Co}$ nm)/Ni($t_{Ni}$ nm)]$_5$/Pt(1 nm) multilayers by combining the four standard magnetic characterization techniques. The magnetic-related hysteresis loops obtained from the field-dependent magnetization $M$ and anomalous Hall resistivity (AHR) $\rho_{xy}$ found that the two serial multilayers with $t_{Co}$ = 0.2 and 0.3 nm have the optimum PMA coefficient $K_U$ well as the highest coercivity $H_C$ at the Ni thickness $t_{Ni}$ = 0.6 nm. Additionally, the magnetic domain structures obtained by Magneto-optic Kerr effect (MOKE) microscopy also significantly depend on the thickness and $K_U$ of the films. Furthermore, the thickness-dependent linewidth of ferromagnetic resonance is inversely proportional to $K_U$ and $H_C$, indicating that inhomogeneous magnetic properties dominate the linewidth. However, the intrinsic Gilbert damping constant determined by a linear fitting of frequency-dependent linewidth does not depend on Ni thickness and $K_U$. Our results could help promote the PMA [Co/Ni] multilayer applications in various spintronic and spin-orbitronic devices.

# 1. INTRODUCTION

Magnetic multilayers with strong perpendicular magnetic anisotropy (PMA) and low magnetic damping have attracted much attention because of their potential applications in high-density magnetic random access memories (MRAM)[1-5] and spin torque nano-oscillators[6-9]. Compared to the in-plane magnetized ferromagnets, ferromagnetic films with PMA facilitate the realization of nonvolatile MRAM with lower power and higher density storage because the latter has lower critical switching current and higher thermal stability than the former as the continuous downscaling of the cell size of devices[10]. In addition, PMA can be an effective magnetic field to achieve zero external magnetic field working spin-torque nano-oscillators with ferromagnets with strong PMA and low damping as its free layer.[11] Therefore, the controllable tailoring PMA of magnetic films is an essential prerequisite for developing high-performance spintronic devices. The magnetic multilayers, e.g., [Co/Pd], [Co/Pt], and [Co/Ni], provide an opportunity to tune their magnetic properties by changing the thickness ratio controllably and the number of bilayer repeats thanks to the interface-induced PMA due to interfacial spin-orbit coupling and interfacial strain relevant magnetoelastic effects[12-17]. Among these PMA multilayers, the PMA [Co/Ni] multilayer also exhibits low damping constant[14], which gets much attention, especially for the fields of current-driven auto-oscillation of magnetization and excitation and manipulation of spin-waves[18, 19]. Furthermore, the PMA [Co/Ni] multilayer also useful for spin-orbit torque devices[20-22]. Therefore, [Co/Ni] multilayers are considered one of the most promising PMA ferromagnets in various spintronic devices. Although there are a few studies on the magnetic anisotropy, magnetotransport, and magnetic damping of Pt/[Co/Ni] multilayers [6, 14, 23], the systematically studied evolution of magnetostatic properties, including the topography of magnetic domains and magnetic dynamics with the thickness ratio of Co and Ni layers for this multilayer film still needs to make a thorough investigation for facilitating it better used in further spintronics.

Here, we systematically investigate how to control the magnetic film PMA by tailoring the interfacial effect by varying the thickness of the Ni layer and its impact on magnetic domain structure and dynamical damping in two serial Co/Ni multilayers with $t_{Co}$ = 0.2 and 0.3 nm. The highest PMA coefficient $K_U \sim 3 \times 10^6$ erg cm$^{-3}$ and coercivity $H_C \sim 250$ Oe are found at the optimum Ni thickness $t_{Ni}$ = 0.6 nm for the studied two serials. The nucleation of the magnetic domain occurs at only a few nucleation sites and gradually expands with magnetic fields for the multilayers with the optimum Ni thicknesses 0.4 nm ~ 0.6 nm. Finally, the intrinsic Gilbert damping constant $\alpha$ is not sensitive to thickness-dependent $K_U$ and domain structures even though the linewidth of ferromagnetic resonance is inversely proportional to $K_U$ and $H_C$, which is dominated by inhomogeneous magnetic properties.

## 2. EXPERIMENT

Two serial multilayers of Pt(5)/[Co(0.2)/Ni($t_{Ni}$)]$_5$/Pt(1) and Pt(5)/[Co(0.3)/Ni($t_{Ni}$)]$_5$/Pt(1), named as Pt/[Co(0.2)/Ni($t_{Ni}$)] and Pt/[Co(0.3)/Ni($t_{Ni}$)] respectively, were deposited on Si/SiO$_2$ substrates at room temperature by dc-magnetron sputtering with Ar pressure $3 \times 10^{-3}$ torr. The unit in parentheses is the thickness in nm. The base pressure of the sputtering deposition chamber is below $2 \times 10^{-8}$ torr. The deposition rate was monitored by the quartz crystal monitor in situ and calibrated by spectroscopic ellipsometry (SE). The static magnetic properties were characterized by the vibrating sample magnetometer (VSM), the anomalous Hall resistivity (AHR) measurement，and the Magneto-optic Kerr effect (MOKE) microscopy, respectively. The films' ferromagnetic resonance (FMR) spectra, obtained by combining coplanar waveguide (CPW) and lock-in techniques, were also adopted to characterize their dynamic magnetic properties. All these magnetic characterizations were performed at room temperature.

## 3. RESULTS AND DISCUSSION

### 3.1 Quasi-static magnetic properties

To directly obtain the thickness dependence of PMA properties in the Co/Ni films, we first performed the magnetic hysteresis loops of samples with different thicknesses using VSM. Figure 1 shows the magnetization hysteresis loops with the out-of-plane and in-plane field geometries for the two serial multilayers of Pt/[Co(0.2)/Ni($t_{Ni}$)] and Pt/[Co(0.3)/Ni($t_{Ni}$)] samples. The well-defined square $M$-$H$ loops under out-of-plane field [Figs.1(a) and (c)] indicate that two studied serial Pt/[Co/Ni] multilayers exhibit a perpendicular magnetic anisotropy. Additionally, the saturation magnetization $M_S$ of the multilayers decreases with increasing the thickness of the Ni layer $t_{Ni}$, from 673 emu cm$^{-3}$ to 495 emu cm$^{-3}$ for Pt/[Co(0.2)/Ni($t_{Ni}$)] and 723 emu cm$^{-3}$ to 639 emu cm$^{-3}$ for Pt/[Co(0.3)/Ni($t_{Ni}$)], which agrees with the much lower $M_S \sim 484$ emu cm$^{-3}$ of the metal nickel compared to that of the cobalt layer $M_S \sim 1\,422$ emu cm$^{-3}$. Based on the out-of-plane and in-plane magnetization hysteresis loops, the perpendicular anisotropy field $H_K$ was determined by using the defined formula for the PMA[12, 24]: $H_K = \frac{2}{M_s}\left(\int_0^{M_S} H_\perp dM - \int_0^{M_S} H_\parallel dM\right) + 4\pi M_S$. The calculated $H_K$, $M_S$ and the coercivity $H_C$, obtained from the $M$-$H$ loops, were summarized below in Fig. 5.

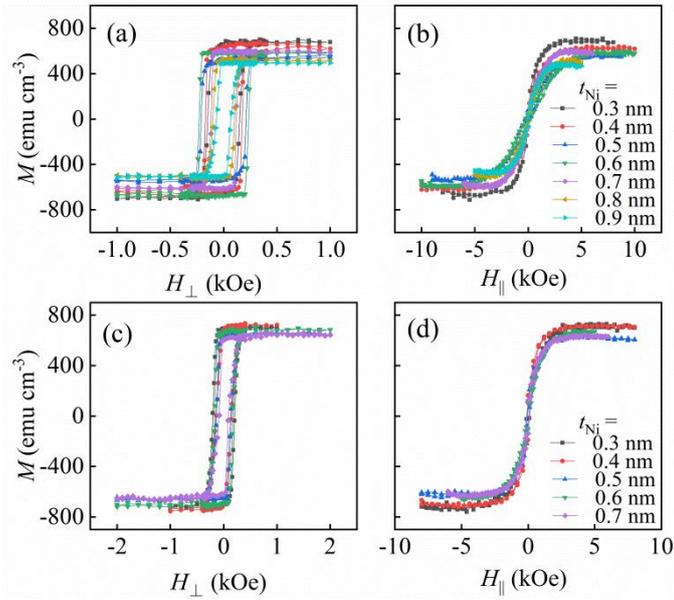

Fig. 1.(a) - (b) Magnetization loops of the films Pt(5)/[Co(0.2)/Ni($t_{Ni}$)]$_5$/Pt(1) with out-of-plane (a) and in-plane (b) magnetic field. (c)-(d) Same as (a)-(b), for Pt(5)/[Co(0.3)/Ni($t_{Ni}$)]$_5$/Pt(1).

These static magnetic properties of the metal Pt/[Co/Ni] multilayer films also can be determined by the electric transports in magnetic field, e.g., anomalous Hall resistivity (AHR) and magnetoresistive effect. Compared to the standard magnetometer above, the electric transports in magnetic field measurements provide an alternative approach and, especially, more useful for spintronic nano-devices because they can easily access the magnetic properties of the microscale and nanoscale samples[25, 26]. Therefore, we also perform the out-of-plane and in-plane AHR loops as a function of the applied magnetic fields for the studied two serial multilayers, as shown in Fig. 2. The coercivity $H_C$ determined from the out-of-plane AHR loops are well consistent with the values obtained by the M-H loops, and are also summarized in the following Fig. 5. Meanwhile, we can calculate $H_K$ of the studied films from the in-plane AHR loops by using the following relation[27]: $H_K = H_\parallel \cdot \tan \arcsin \left( \frac{\rho_{xy(H)}}{\rho_{xy(0)}} \right) + 4\pi M_S$, where $\rho_{xy}(0)$ is the AHR value at zero in-plane field. The evolution of $H_K$ with the thickness of the Ni layer is overall consistent with the results determined by the VSM measurement. Furthermore, the AHR measurements also provide us the additional information, which can not be easily accessed by VSM, about the studied two serial Pt/[Co/Ni] multilayers. For example, we find that the in-plane AHR near-zero magnetic field is much smaller than the out-of-plane AHR for the samples with certain Ni thickness, indicating that these samples form the multi-domain structures at the low in-plane magnetic fields. Therefore, the value of the difference between out-of-plane and in-plane AHR at near-zero fields hints that the different Ni thickness films may exhibit distinct magnetic domain structures[28].

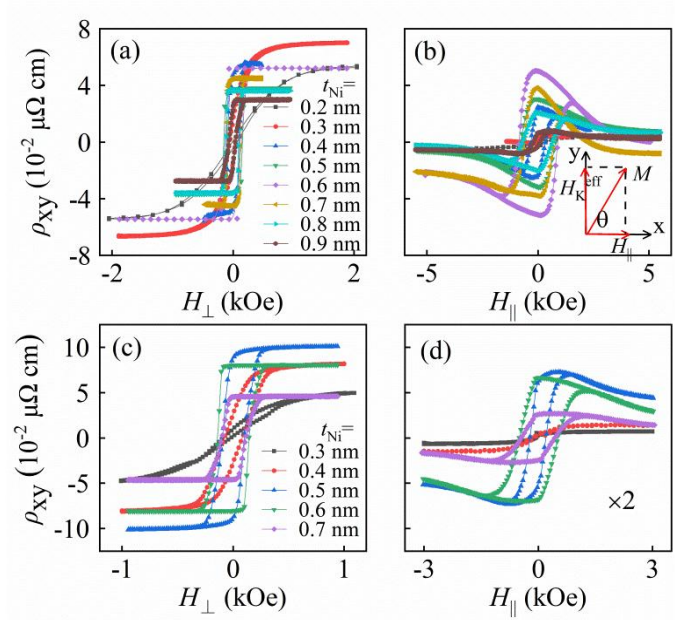

Fig. 2.(a)-(b) Anomalous Hall resistivity as a function of out-of-plane (a) and in-plane magnetic field with 5° tilt angle from the file plane (b) for the samples Pt(5)/[Co(0.2)/Ni($t_{Ni}$)]$_5$/Pt(1). The inset shows the geometric relationship between magnetic field, magnetization and effective anisotropic field $H_K^{eff} = H_K - 4\pi M_S$ (c)-(d) Same as (a)-(b), for Pt(5)/[Co(0.3)/Ni($t_{Ni}$)]$_5$/Pt(1). All AHR were measured by using the films patterned into a 0.3 × 10 mm Hall cross.

**3.2 Magnetic domain structures**

To directly reveal the evolution of magnetic domain structure with Ni thickness and the detail of magnetization reversal process under external field, the MOKE microscopy is also performed for Pt/[Co(0.2)/Ni($t_{Ni}$)] serial samples. Figure 3 shows MOKE loops and the representative MOKE images during scanning out-of-plane magnetic fields. Bright and dark regions represent the domains with magnetic moment point-up and point-down, respectively. Except for the multilayer with the nickel thickness $t_{Ni}$ = 0.2 nm [Fig. 3(a)], all films exhibit a well-defined PMA corresponding rectangular or dumbbell-shape hysteresis loops [Figs. 3(b)-(h)]. The MOKE images show that the sample with $t_{Ni}$ = 0.3 nm begins to nucleate with numerous nucleation points approximately uniform dispersion on the whole film at the field of 138 Oe, then gradually expand with increasing field, and saturate to the uniform state at $H_\perp$ > 300 Oe [Fig.3(b)]. Combining with field-dependent magnetic susceptibility and AHR characterizations, $H_C$ and $H_K$ of Pt/[Co(0.2)/Ni($t_{Ni}$)] first enhance with increasing $t_{Ni}$, reach the maximum at $t_{Ni}$ = 0.6 nm, and then reduce with continue increasing $t_{Ni}$ to 0.9 nm. For the samples with $0.4\,\text{nm} \leq t_{Ni} \leq 0.6\,\text{nm}$, the MOKE loops exhibit a well-defined rectangular shape. Meanwhile, in contrast to $t_{Ni}$ = 0.3 nm, only a few nucleation points appear at the critical magnetic field, and then begin to expand the magnetic domain with a continuously increasing field. For increasing $t_{Ni}$ to above 0.7 nm, the MOKE loops begin to transfer into the dumbbell shape from the prior well-defined rectangular shape. The corresponding MOKE images show that the films with $t_{Ni} \geq 0.7$ nm form tree-like domain walls with more irregular branchings as increasing $t_{Ni}$ during the field range near below the saturation field [the middle picture of Fig.3]. In other words, for the films with $t_{Ni} \geq 0.7$ nm, the length of the domain wall increases with increasing $t_{Ni}$, which is consistent with the trend of dependence of $H_K$ on Ni thickness. As we know that the total energy density of the magnetic domain wall per unit area $\gamma_w$ is proportional to magnetic anisotropy energy, exchange energy and demagnetization energy based on the widely recognized formula[29]: $\gamma_w = A\frac{\pi^2}{\delta} + \frac{K_U}{2}\delta + \frac{M_S^2 \mu_0 \delta^2}{4}\frac{1}{t+\delta}$, where A is the exchange constant, δ is the thickness of the domain wall, $K_U$ is the PMA coefficient, $M_S$ is the saturation magnetization, $t$ is the thickness of the entire film, and $\mu_0$ is the permeability of vacuum. To minimize the total energy of the entire film, the volume (or length) of the domain wall needs to reduce and increase demagnetization energy when $K_U$ increases with increasing $t_{Ni}$ in the range of 0.4 nm ~ 0.6 nm.

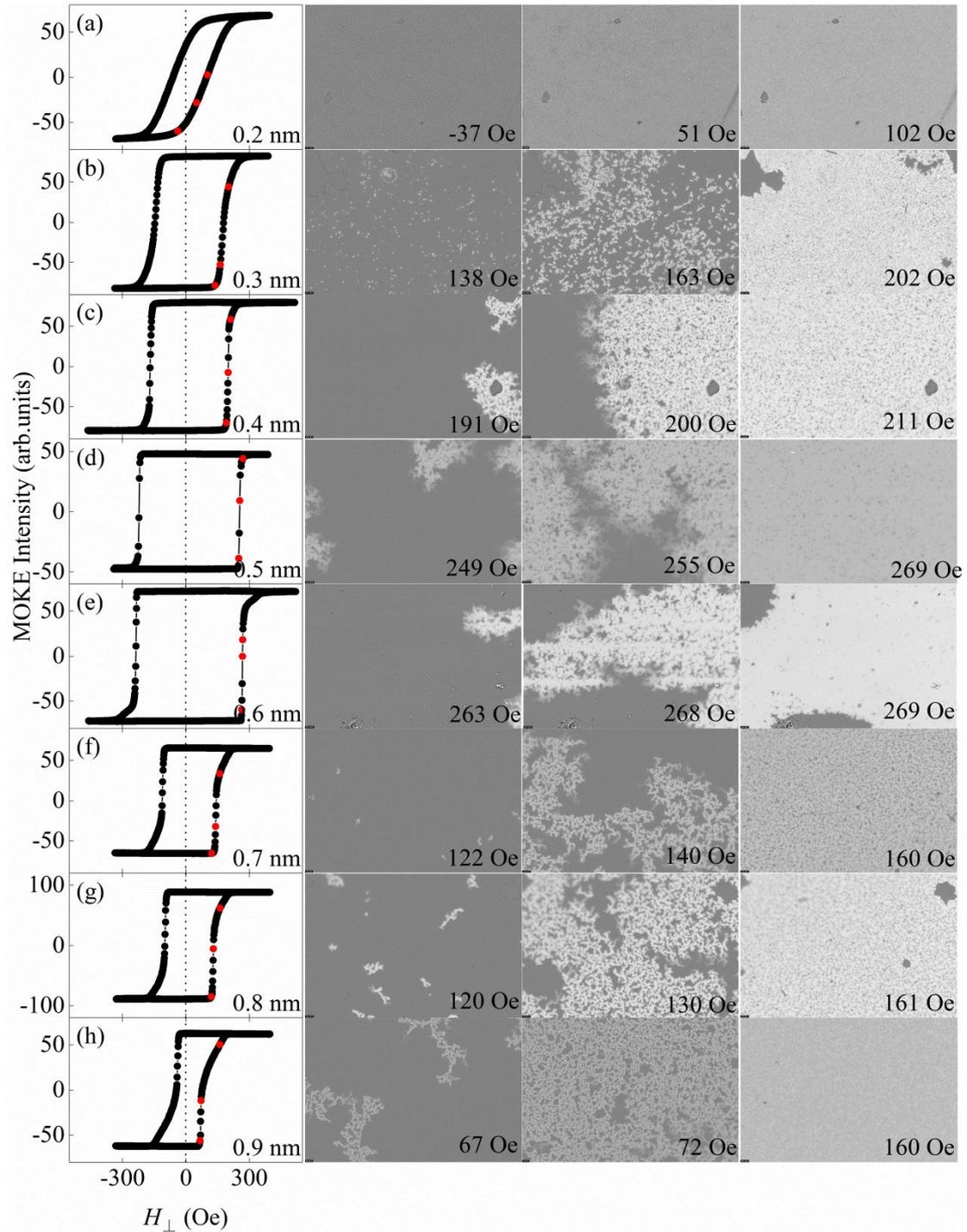

Fig. 3.(a)-(h) Magneto-optic Kerr hysteresis loops and magnetic domain images of the films Pt(5)/[Co(0.2)/Ni($t_{Ni}$)]$_5$/Pt(1) with labeled thickness $t_{Ni}$ = 0.2 nm (a), 0.3 nm (b), 0.4 nm (c), 0.5 nm (d), 0.6 nm (e), 0.7 nm (f), 0.8 nm (g), 0.9 nm (h), respectively. The corresponding magnetic domain images with the size of 100 μm × 150 μm

were obtained at the labeled out-of-plane magnetic fields (also marked as the red dots on loops).

**3.3 Magnetization dynamics**

To further investigate the Ni thickness-dependent magnetization dynamics of Co/Ni multilayers, we perform the broadband FMR measurement with the external field perpendicular to the film plane. All FMR measurements were carried out with a home-made differential FMR measurement system combining Lock-in technique at room temperature. A continuous-wave Oersted field with a selected radio frequency is generated via connecting coplane waveguide (CPW) with an RF generator, which produces a microwave signal to excite FMR of ferromagnetic film, which with film surface was adhered on the CPW. The RF power used in the experiments is 15 dBm. To improve the signal-to-noise ratio (SNR), a lock-in detection technique is employed through the modulation of signals. The modulation of a direct current (DC) magnetic field $H$ is provided by a pair of secondary Helmholtz coils powered by an alternating current (AC) source with 129.9 Hz [see Fig. 4(a)][16, 30]. The differential absorption signal is measured by sweeping the magnetic field with a fixed microwave frequency. The representative FMR spectrum of Pt(5)/[Co(0.2)/Ni(0.3)]$_5$/Pt(1) obtained at 9 GHz is shown in the inset of Fig. 4(b). The differential FMR spectrum can be well fitted by using a combination of symmetric and antisymmetric Lorentzian function, as follows :

$$\frac{dP}{dH} = V_S \frac{4\Delta H(H-H_{res})}{[4(H-H_{res})^2+(\Delta H)^2]^2} + V_A \frac{(\Delta H)^2-4(H-H_{res})^2}{[4(H-H_{res})^2+(\Delta H)^2]^2} , \qquad (1)$$

where $V_S$ and $V_A$ represent the symmetric and antisymmetric factors, $H$ is the external magnetic field, $H_{res}$ is the resonance field, and $\Delta H$ is the linewidth of FMR corresponding $\sqrt{3}$ times of the peak-to-dip width in the FMR spectrum. The relationship between the frequency $f$ and the resonance field $H_{res}$ of the two series of Pt/[Co(0.2)/Ni($t_{Ni}$)] and Pt/[Co(0.3)/Ni($t_{Ni}$)] samples [Figs. 4(b) and (d)] can be well fitted by the Kittel equation [31]

$$f = \left(\frac{\gamma}{2\pi}\right)(H_{\text{res}} + H_{\text{eff}}), \tag{2}$$

where $\frac{\gamma}{2\pi} = 2.8 \text{ MHz Oe}^{-1}$ is the gyromagnetic ratio, $H_{\text{eff}}$ is effective demagnetization[32] $H_{\text{eff}} = H_K - 4\pi M_S$. Therefore, the magnetic anisotropy field $H_K$ also can be directly determined from the dispersion relation of $f$ versus $H_{\text{res}}$ by using a parameter $M_S$ obtained by VSM. In addition, we can obtain the intrinsic Gilbert damping α by fitting the experimental data of linewidth $\Delta H$ versus resonance frequency [Figs.4(c) and (e)] with the formula: $\Delta H = \Delta H_0 + \frac{4\pi\alpha f}{\gamma}$, where $\Delta H_0$ is an inhomogeneous linewidth independent of the frequency, and the second term is the intrinsic linewidth linearly proportional to the frequency. The inhomogeneous linewidth of samples is derived from roughness, defects and inhomogeneous PMA and magnetization[33].

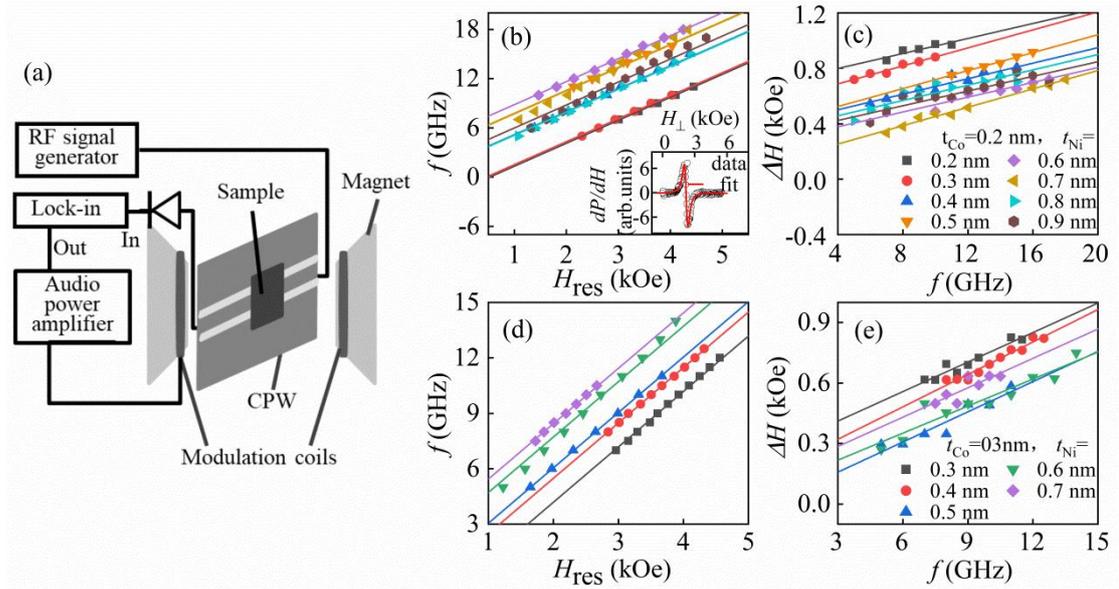

Fig. 4.(a) Differential FMR spectra experimental setup. (b) Dependence of the resonance field $H_{\text{res}}$ on the frequency $f$ with the out-of-plane field for the films Pt(5)/[Co(0.2)/Ni($t_{\text{Ni}}$)]$_5$/Pt(1). Solid lines indicate the Kittel fitting curves. The inset is the representative FMR spectrum obtained at 9 GHz, which can be well fitted by

Eq.(1) (solid red line). (c) The linewidth versus frequency (symbols) for the samples Pt(5)/[Co(0.2)/Ni($t_{Ni}$)]$_5$/Pt(1). The solid line is a linear fitting, which can extract the corresponding damping constant α based on Eq.(2). (d)-(e) Same as (b)-(c), for the films Pt(5)/[Co(0.3)/Ni($t_{Ni}$)]$_5$/Pt(1).

Figure 5 summarizes the dependence of the determined material parameters: the saturation magnetization $M_S$, the coercivity $H_C$, the anisotropy field $H_K$, the inhomogeneous linewidth $\Delta H_0$ and the magnetic damping constant α on Ni thickness $t_{Ni}$ for the studied two series of Pt/[Co(0.2)/Ni($t_{Ni}$)] and Pt/[Co(0.3)/Ni($t_{Ni}$)] samples. The determined $H_K$ by three independent methods shows an overall consistent behavior. The $H_K$ begins to increase with increasing $t_{Ni}$, and reaches the maximum at $t_{Ni}$ ~ 0.6 nm, whereafter reduces again with continuing to increase $t_{Ni}$. Several reasons account for this phenomenon. First, the magnetic anisotropy of the studied multilayer is mainly contributed from the interfacial magnetic anisotropy of the Co/Ni and Pt/Co interfaces[34]. Second, the Co/Ni multilayers' interface quality depends highly on the Ni layer's thickness. In other words, too thin nickel layer may not get a good Co/Ni interface due to inevitable elements diffusion during sputtering deposition. However, the $H_K$ will drop due to reducing the ratio of the interfacial anisotropy to the volume anisotropy energy if the Ni layer is too thick. Like $H_K$, the $H_C$ shows a similar trend with varying thickness of the Ni layer. As we well know that the coercivity depends on PMA, as well as defects-induced pinning effects. But, in our case, the results show that the combination of PMA and magnetization-relevant demagnetization field dominate the coercivity, which can be well explained by the Brown formula[35]: $H_C = \frac{2K_U}{M_S} - NM_S$, where $K_U = (M_S * H_K)/2$ and $N$ are the magnetic anisotropy constant and the demagnetization factor of the film, respectively.

Figures 5(d) and 5(i) show the inhomogeneous linewidth ($\Delta H_0$) of FMR spectra as a function of $t_{Ni}$ for Pt/[Co(0.2)/Ni($t_{Ni}$)] and Pt/[Co(0.3)/Ni($t_{Ni}$)], respectively. For thin thickness Ni samples, island structures are most likely formed. This results in a broadening of the resonance linewidth due to a distribution of effective internal anisotropy and demagnetization fields[37]. One can see that the minimum linewidth of two serial samples corresponds to the maximum PMA field $H_K$, suggesting the inhomogeneous magnetic anisotropy-induced linear broadening is the minimum at the optimum PMA condition[36]. Although the intrinsic damping constant is almost independent of the Ni thickness for the studied two serials, but the Pt/[Co(0.2)/Ni($t_{Ni}$)] films have a lower damping constant α ~ 0.04 than α ~ 0.07 of Pt/[Co(0.3)/Ni($t_{Ni}$)]. The obvious difference in damping constant between the two serial multilayer systems indicates that the former has better magnetic dynamic properties.

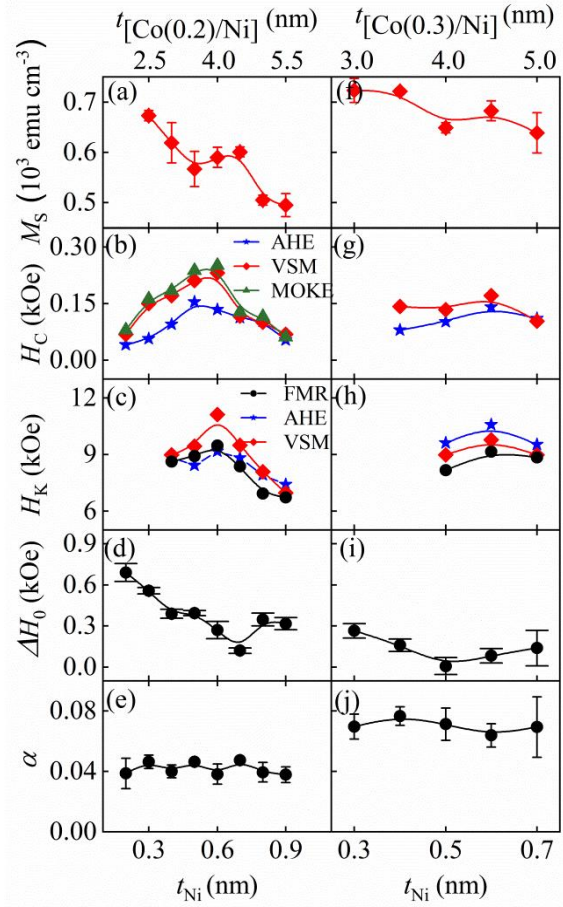

Fig. 5.(a)-(e) Dependence of the saturation magnetization $M_S$ (a), the coercivity $H_C$ (b), the anisotropy field $H_K$ (c), the inhomogeneous linewidth $\Delta H_0$ (d) and the magnetic damping constant α (e) on the Ni thickness $t_{Ni}$ in the films Pt(5)/[Co(0.2)/Ni($t_{Ni}$)]$_5$/Pt(1). (f)-(j) Same as (a)-(e) for the samples Pt(5)/[Co(0.3)/Ni($t_{Ni}$)]$_5$/Pt(1). $M_S$, $H_C$, and $H_K$ were determined from the previous

magnetization loops, AHR loops, MOKE loops, and the ferromagnet resonance spectra. The linewidth was determined by fitting the experimental FMR spectrum with a Lorentzian function based on Eq.(1). The magnetic damping constant was obtained by a linear fitting of $\Delta H$ vs. $f$ curves based on Eq.(2).

## 4. CONCLUSION

Ni thickness effect on the static magnetic properties and magnetic dynamics of Pt(5 nm)/[Co(0.2 nm and 0.3 nm)/Ni($t_{Ni}$ nm)]$_5$/Pt(1 nm) multilayers demonstrate that the two studied serial multilayer systems exhibit the optimum PMA coefficient $K_U$ well as the highest coercivity $H_C$ at the Ni thickness $t_{Ni}$ = 0.6 nm. The MOKE images further confirm that the maximum $K_U$ corresponds to the magnetic domain structure with the shortest length of domain wall through minimizing the total energy, which consists of magnetic anisotropy energy, exchange energy, and demagnetization energy. Furthermore, the frequency-dependent FMR spectra show that the damping constant remains almost constant with the different Ni thicknesses for both serials, but the Pt/[Co(0.2)/Ni($t_{Ni}$)] multilayer serial has a lower damping constant α ~ 0.04 than 0.07 of the Pt/[Co(0.3)/Ni($t_{Ni}$)] serial. According to the obtained results, we find that the optimum PMA coefficient $K_U$ = 3.3 × 10$^6$ erg cm$^{-3}$, the highest coercivity $H_C$ = 250 Oe, and as well as the lowest damping constant α = 0.04 can be achieved at Pt(5)/[Co(0.2)/Ni(0.6)]$_5$/Pt(1). Our results of optimizing magnetic properties of the Pt/[Co/Ni] multilayer by tuning the ratio of Co/Ni layers is helpful to facilitate its applications in various spintronic devices.

**Acknowledgement**： Project is supported by the National Natural Science Foundation of China (Grant Nos. 11774150, 12074178, 12004171, 12074189 and 51971109), the Applied Basic

Research Programs of Science and Technology Commission Foundation of Jiangsu Province, China (Grant No. BK20170627), the National Key R&D Program of China (Grant No. 2018YFA0209002), the Open Research Fund of Jiangsu Provincial Key Laboratory for Nanotechnology, and the Scientific Foundation of Nanjing University of Posts and Telecommunications (NUPTSF) (Grant No. NY220164).

Corresponding authors, Email: chenlina@njupt.edu.cn, rhliu@nju.edu.cn

# References

[1] Li Q Y, Zhang P H, Li H T, Chen L N, Zhou K Y, Yan C J, Li L Y, Xu Y B, Zhang W X, Liu B, Meng H, Liu R H and Du Y W 2021 *Chin. Phys. B* **30** 047504

[2] Yang M S, Fang L and Chi Y Q 2018 *Chin. Phys. B* **27** 098504

[3] Zhu T 2014 *Chin. Phys. B* **23** 047504

[4] Su Y C, Lei H Y and Hu J G 2015 *Chin. Phys. B* **24** 097506

[5] Fu Q W, Zhou K Y, Chen L N, Xu Y B, Zhou T J, Wang D H, Chi K Q, Meng H, Liu B, Liu R H and Du Y W 2020 *Chin.Phys.Lett.* **37** 117501

[6] Liu R H, Lim W L and Urazhdin S 2015 *Phys. Rev. Lett.* **114** 137201

[7] Chen L N, Gao Z Y, Zhou K Y, Du Y W and Liu R H 2021 *Phys.Rev.Appl.* **16** 034044

[8] Chen L N, Chen Y, Zhou K Y, Li H T, Pu Y, Xu Y B, Du Y W and Liu R H 2021 *Nanoscale.* **13** 7838

[9] Li L Y, Chen L N, Liu R H and Du Y W 2020 *Chin. Phys. B* **29** 117102

[10] Wang G Z, Zhang Z Z, Ma B and Jin Q Y 2013 *J. Appl. Phys.* **113** 17C111

[11] Guo Y Y, Zhao F F, Xue H B and Liu Z J 2016 *Chin.Phys.Lett.* **33** 037501

[12] Andrieu S, Hauet T, Gottwald M, Rajanikanth A, Calmels L, Bataille A M, Montaigne F, Mangin S, Otero E, Ohresser P, Le Fèvre P, Bertran F, Resta A, Vlad A, Coati A and Garreau Y 2018 *Phys. Rev. Materials* **2** 064410

[13] Li R Z, Li Y C, Sheng Y and Wang K Y 2021 *Chin. Phys. B* **30** 028506

[14] Song H S, Lee K D, Sohn J W, Yang S H, Parkin S S P, You C Y and Shin S C 2013 *Appl. Phys. Lett.* **102** 102401

[15] Nakazawa S, Obinata A, Chiba D and Ueno K 2017 *Appl. Phys. Lett.* **110** 062406

[16] Li Q Y, Xiong S Q, Chen L N, Zhou K Y, Xiang R X, Li H T, Gao Z Y, Liu R H and Du Y W 2021 *Chin.Phys.Lett.* **38** 047501

[17] Huang L A, Wang M Y, Wang P, Yuan Y, Liu R B, Liu T Y, Lu Y, Chen J R, Wei L J, Zhang W, You B, Xu Q Y and Du J 2022 *Chin. Phys. B* **31** 027506

[18] Chen L N, Urazhdin S, Zhou K Y, Du Y W and Liu R H 2020 *Phys.Rev.Appl.* **13** 024034

[19] Chen L N, Gu Y Y, Zhou K Y, Li Z S, Li L Y, Gao Z Y, Du Y W and Liu R H 2021 *Phys. Rev. B* **103** 144426

[20] Cai K M, Yang M Y, Ju H L, Wang S M, Ji Y, Li B H, Edmonds K W, Sheng Y, Zhang B, Zhang N, Liu S, Zheng H Z and Wang K Y 2017 *Nat. Mater.* **16** 712

[21] Yang M Y, Cai K M, Ju H L, Edmonds K W, Yang G, Liu S, Li B H, Zhang B, Sheng Y, Wang S G, Ji Y and Wang K Y 2016 *Sci. Rep.* **6** 20778

[22] Yang M Y, Deng Y C, Wu Z H, Cai K M, Edmonds K W, Li Y C, Sheng Y, Wang S M, Cui Y, Luo J, Ji Y, Zheng H Z and Wang K Y 2019 *IEEE Electron Device Lett* **40** 1554

[23] Cui Y W, Feng X Y, Zhang Q H, Zhou H A, Jiang W J, Cao J W, Xue D S and Fan X L 2021 *Phys. Rev. B* **103** 024415

[24] Bai Q W, Guo B, Yin Q and Wang Sh Y 2022 *Chin. Phys. B* **31** 017501

[25] Zhang P, Xie K X, Lin W W, Wu D and Sang H 2014 *Appl. Phys. Lett.* **104** 082404

[26] Lohmann M, Su T, Niu B, Hou Y S, Alghamdi M, Aldosary M, Xing W Y, Zhong J N, Jia S, Han W, Wu R Q, Cui Y T and Shi J 2019 *Nano Lett.* **19** 2397

[27] Chiba D, Kawaguchi M, Fukami S, Ishiwata N, Shimamura K, Kobayashi K and Ono T 2012 *Nat Commu.* **3** 888

[28] Tang J X, Xu G Z, You Y R, Xu Z, Zhang Z, Chen X, Gong Y Y and Xu F 2020 *Appl. Phys. Lett.* **117** 202402

[29] Hubert A and Schäfer R 1988 *magnetic domains* (World Publishing Corporation) p.148

[30] Fu Q W, Li Y, Chen L N, Ma F S, Li H T, Xu Y B, Liu B, Liu R H and Du Y W 2020 *Chin.Phys.Lett.* **37** 087503

[31] Nembach H T, Silva T J, Shaw J M, Schneider M L, Carey M J, Maat S and Childress J R 2011 *Phys. Rev. B* **84** 054424

[32] Mao S W, Lu J, Yang L, Ruan X Z, Wang H L, Wei D H, Xu Y B and Z J H 2020 *Chin.Phys.Lett.* **37** 058501

[33] Xie H k, Pan L N, Cheng X H, Zhu Z T, Feng H M, Wang J B and Liu Q F 2018 *J. Magn. Magn. Mater.* **461** 19

[34] Arora M, Hübner R, Suess D, Heinrich B and Girt E 2017 *Phys. Rev. B* **96** 024401

[35] Kronmüller H, Durst K -D, Sagawa M 1988 *J. Magn. Magn. Mater.* **74** 291

[36] Shaw J M, Nembach H T and Silva T J 2010 *J. Appl. Phys.* **108** 093922

[37] Platow W, Anisimov A N, Dunifer G L, Farle M and Baberschke K 1998 *Phys. Rev. B* **58** 561